\documentstyle[preprint,aps]{revtex}

\begin{document}

\draft

\preprint{LBL-34664}

\title{
Calculating Dilepton Rates from Monte Carlo
Simulations of Parton Production\thanks{\baselineskip=12pt
This work was supported by the Director, Office of Energy
Research, Division of Nuclear Physics of the Office of High
Energy and Nuclear Physics of the U.S. Department of Energy
under Contract No. DE-AC03-76SF00098. }}
\author{K. J. Eskola$^{1,2}$ and Xin-Nian Wang$^1$}
\address{
$^1$Nuclear Science Division, Mailstop 70A-3307,
        Lawrence Berkeley Laboratory\\
University of California, Berkeley, California 94720.\\
$^2$Laboratory of High Energy Physics, P.O. Box 9,
        SF-00014 University of Helsinki, Finland.\cite{address}}
\date{Oct. 15, 1993}
\maketitle
\begin{abstract}

  To calculate dilepton rates in a Monte Carlo simulation
of ultrarelativistic heavy ion collisions, one usually
scales the number of similar QCD processes by
a ratio of the corresponding differential probabilities.
We derive the formula for such a
ratio especially for dilepton bremsstrahlung processes.
We also discuss the non-triviality of including higher
order corrections to direct Drell-Yan process.
The resultant mass spectra from our Monte Carlo
simulation are consistent with the semi-analytical
calculation using dilepton fragmentation functions.

\end{abstract}
\narrowtext

\section{INTRODUCTION}

In a previous paper\cite{KEXW}, we investigated
dilepton production associated with minijet final
state radiation in heavy ion collisions at collider
energies, using dilepton fragmentation functions which
can be evaluated perturbatively. The dilepton pairs
from the fragmentation of minijets are found to be
comparable to direct Drell-Yan (DY)
for small invariant mass $M\sim$ 1--2 GeV/c$^2$ at the
highest energy  of the Brookhaven Relativistic Heavy Ion
Collider (RHIC). At the CERN Large Hadron Collider (LHC)
energies, the associated dilepton production becomes
dominant over a relative large range of the invariant mass.

Due to the relatively large invariant dilepton mass,
$M\gg \Lambda$, the radiative corrections are calculable
in pQCD up to all orders in the leading logarithm
approximation. Collinear approximation is also used
in convoluting the obtained dilepton fragmentation
functions with minijet cross sections to compute
the radiative contributions to dilepton production.
Since there exist Monte Carlo simulations of QCD
cascading \cite{ODOR,WEBB,PYTH} which can take
into account many other effects,
like multiple ladder structure, it is important
to check our semi-analytical approach with
realistic Monte Carlo simulations. In this way,
we can address the validity of the approximations
we made in the fragmentation function approach \cite{KEXW}.

To directly simulate dilepton production in a Monte Carlo
event generator is rather difficult due to the small QED
coupling constant as compared to that of QCD. To overcome
this difficulty, one can multiply the number of specific
QCD processes,  which resemble those of dilepton production
by an appropriate ratio of the corresponding
differential probabilities, as has been tried
by Geiger and Kapusta \cite{KGJK}. However, the problem
is more complicated than one might first think. For radiative
dilepton production, one has to take into account the
fact that the corresponding radiated quarks and
gluons in QCD can have
further bremsstrahlung which is
different from the QED case. The  additional
bremsstrahlung gives rise to an extra Sudakov form factor
and one must include it in the differential ratio
to give the correct dilepton emission. We will derive a
formula for the differential ratio and demonstrate that the
resultant simulation is consistent with our previous
semi-analytical calculation in Ref.~\cite{KEXW}.

The problem of simulating direct DY process among
QCD processes lies in how to take into account higher
order, ${\cal O}(\alpha^2\alpha_s)$, corrections. We will
discuss when  an overall
$K$ factor, as used in most Monte Carlo simulations
of QCD hard processes in hadronic and nuclear collisions,
is sufficient enough to simulate the
QCD contributions. We also consider how
possible double counting can be avoided.

The remainder of the paper is organized as follows.
In the next Section, we will discuss how to calculate
the differential ratio to obtain dilepton emission
from a Monte Carlo simulation of the corresponding
pQCD processes. In Section III, we will discuss how
to simulate direct DY processes, especially how to
take into account higher order corrections. The
results of our simulation will be compared to
semi-analytical calculations in Section IV,
and finally, a summary with some discussions
is given in Section V.

\section{SIMULATION OF DILEPTON BREMSSTRAHLUNG}

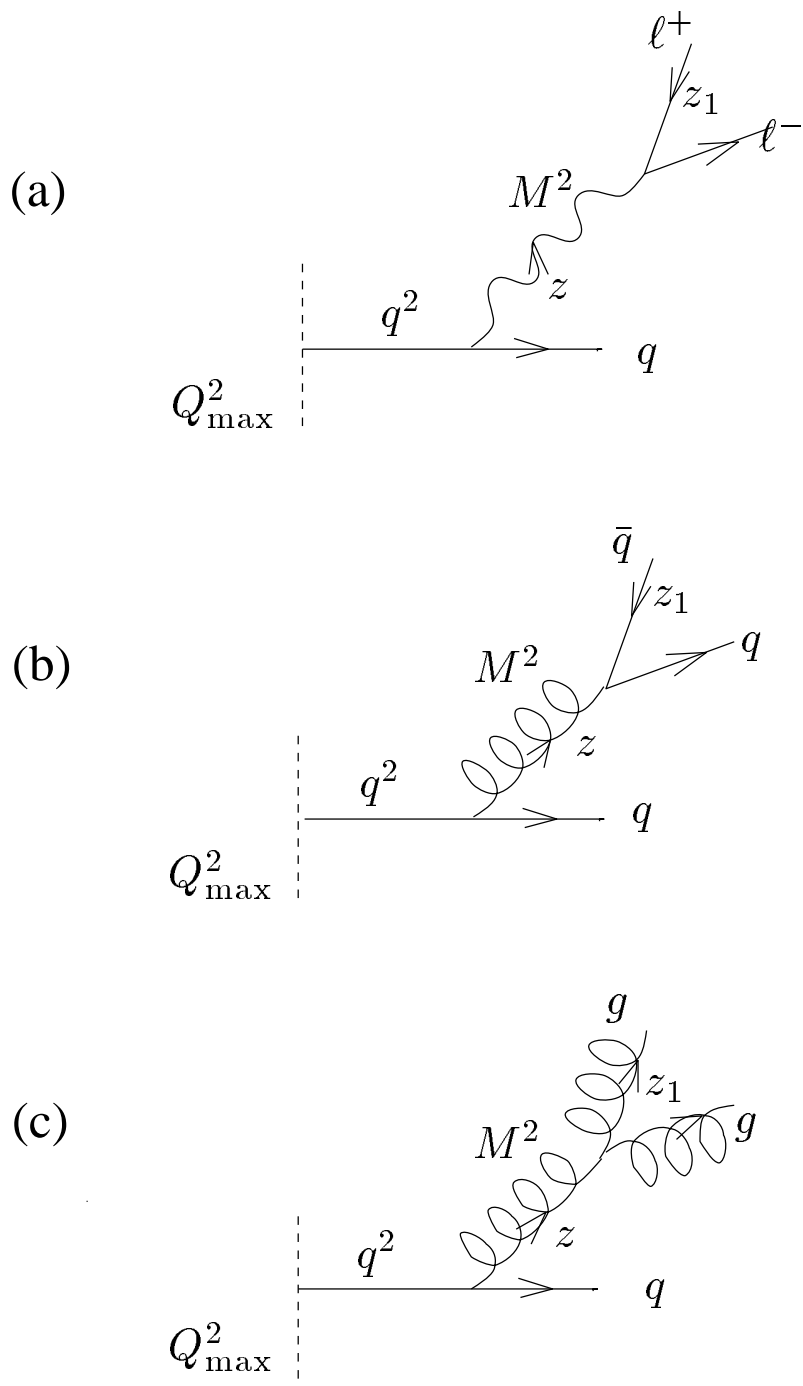
\begin{figure}
\caption{Illustration of (a) dilepton, (b) quark-antiquark
        pair, and (c) two-gluon emission from a quark line.
        The dashed lines present the preceding radiation or
        scattering processes which kinematically determines
        the maximum value, $Q^2_{\rm max}$, of the quark
        virtuality $q^2$.}
\label{fig1}
\end{figure}

If a Monte Carlo generator does
not have the QED processes built in, one can
calculate the absolute cross sections of the QED processes by
scaling the differential cross sections of certain types of
QCD processes. The scaling factor however depends on both the
QED and QCD processes. For
dilepton production through bremsstrhlung, one would naively think
that a scaling factor between virtual photon and virtual
gluon radiation processes from a quark line is enough.
However, the probability to find a virtual gluon
with fixed invariant mass depends on the probability that
the gluon does not have any further radiations to degrade its
virtuality. One therefore should use the scaling factor
between process (a) and processes (b) and (c) in Fig.~\ref{fig1}.

        The Monte Carlo simulation of QCD cascading is carried
out by giving for each vertex of the radiation tree, such as
those in Fig.~\ref{fig1}(b), a {\it normalized} probability distribution.
Given the maximum virtuality $Q^2_{\rm max}$ of a particular
process, the normalized probability for the off-shell
parton $a$ with $q^2\le Q^2_{\rm max}$ to branch into
partons $b$ and $c$ of light-cone momentum fractions
$z$ and $1-z$ is \cite{ODOR,WEBB},
\begin{equation}
        d{\cal P}_{a\rightarrow bc}(q^2,z)= \frac{dq^2}{q^2}
        dz\,P_{a\rightarrow bc}(z)\frac{\alpha_s[z(1-z)q^2]}{2\pi}
        \frac{{\cal S}_a(Q^2_{\rm max})}{{\cal S}_a(q^2)}.
                 \label{eq:shr1}
\end{equation}
The relative transverse momentum between the radiated partons
$b$ and $c$ is given by,
\begin{equation}
        q_T^2=z(1-z)\left(q^2-\frac{q_b^2}{z}-\frac{q_c^2}{1-z}
        \right). \label{eq:shr2}
\end{equation}
Note that the variable $z(1-z)q^2$ in the strong coupling
constant in Eq.~\ref{eq:shr1} is approximately $q_T^2$.
In a Monte Carlo simulation, the time-like
branching is usually
terminated at $q^2\le\mu^2_0$, where the physics of
nonperturbative hadronization sets in. By requiring
$q^2_b, q^2_c\geq\mu_0^2$ and the relative transverse
momentum $q_T$ to be real, one defines
the kinematically allowed region of the phase space as
\begin{eqnarray}
        4\mu_0^2< & q^2 & <Q^2_{\rm max}; \nonumber \\
        \epsilon(q)< & z & <1-\epsilon(q),\ \
        \epsilon(q)=\frac{1}{2}(1-\sqrt{1-4\mu_0^2/q^2}).
                        \label{eq:eps}
\end{eqnarray}
If the virtuality of one of the radiated partons is
fixed to $q^2_b=M^2$, the above region of the phase
space is modified to,
\begin{eqnarray}
        (\mu_0+M)^2< & q^2 & <Q^2_{\rm max}; \nonumber \\
        \epsilon_1(q,M)< & z & <\epsilon_2(q,M),
\end{eqnarray}
where,
\begin{equation}
        \epsilon_{1,2}(q,M)=\frac{1}{2}\left[1+
        \frac{M^2-\mu_0^2}{q^2}\mp\sqrt{
        \left(1+\frac{M^2-\mu_0^2}{q^2}\right)^2
        -4\frac{M^2}{q^2}}\right]. \label{eq:e12}
\end{equation}
In Eq.~\ref{eq:shr1}, the Sudakov form factor
${\cal S}_a(q^2)$ is defined as
\begin{equation}
        {\cal S}_a(q^2)=\exp\left\{
        -\int_{4\mu_0^2}^{q^2}\frac{dk^2}{k^2}
         \int_{\epsilon(k)}^{1-\epsilon(k)}dz\sum_{b,c}
        P_{a\rightarrow bc}(z)\frac{\alpha_s[z(1-z)k^2]}{2\pi}
        \right\}, \label{eq:sdk}
\end{equation}
so that ${\cal S}_a(Q^2_{\rm max})/{\cal S}_a(q^2)$ is
the probability for parton $a$ not to have any branching
between $Q^2_{\rm max}$ and $q^2$. The contribution of
QED processes to the Sudakov form factor is negligible.

        Based on the above probability distributions, we can
write down the differential probability for a quark to radiate
{\em either} a $q\bar{q}$ pair {\em or} two gluons via
an intermediate virtual gluon with invariant mass $M$,
as shown in Figs.~\ref{fig1}(a) and (b),
\begin{eqnarray}
        \frac{d{\cal P}_{q\rightarrow {\rm all}}}{dM^2}&=&
        \frac{1}{M^2}
        \int_{(M+\mu_0)^2}^{Q^2_{\rm max}}\frac{dq^2}{q^2}
        \int_{\epsilon_1(q,M)}^{\epsilon_2(q,M)}dz
        P_{q\rightarrow gq}(z)\frac{\alpha_s[z(1-z)q^2]}{2\pi}
        \frac{{\cal S}_q(Q^2_{\rm max})}{{\cal S}_q(q^2)} \nonumber \\
        & &\int_{\epsilon(M)}^{1-\epsilon(M)}dz_1
        [n_fP_{g\rightarrow q\bar{q}}(z_1)+P_{g\rightarrow gg}(z_1)]
        \frac{\alpha_s[z_1(1-z_1)M^2]}{2\pi}
        \frac{{\cal S}_g((q-\mu_0)^2)}{{\cal S}_g(M^2)}.
        \label{eq:ratio}
\end{eqnarray}
Notice the variables in the second set of Sudakov form factors.
Since $q^2$ is the actual virtuality of the quark line preceding
the gluon radiation and the daughter quark line has at least
virtuality of $\mu_0^2$, the maximum value of the gluon virtuality,
$M^2$, is then $(q-\mu_0)^2$. Due to the same reason, the lower
limit of the integration over $q^2$ is $(M+\mu_0)^2$.

        Similarly, the differential branching probability
for the dilepton production via diagram (a) in Fig.~\ref{fig1} is,
\begin{eqnarray}
        \frac{d{\cal P}_{q\rightarrow {\rm DL}}}{dM^2}&=&
        \frac{e_q^2}{M^2}\int_{(M+\mu_0)^2}^{Q^2_{\rm max}}
        \frac{dq^2}{q^2}\int_{\epsilon_1(q,M)}^{\epsilon_2(q,M)}dz
        P_{q\rightarrow \gamma q}(z)\frac{\alpha}{2\pi}
        \frac{{\cal S}_q(Q^2_{\rm max})}{{\cal S}_q(q^2)} \nonumber \\
        & &\int_0^1dz_1 P_{\gamma\rightarrow \ell^+\ell^-}(z_1)
        \frac{\alpha}{2\pi},
\end{eqnarray}
where again we neglect the QED contribution to the Sudakov
form factor, and the integration over $z$ and $z_1$ can
be carried out analytically using Eq.~\ref{eq:e12}. Notice
that the differential probability for the dilepton bremsstrahlung
has one Sudakov form factor less than the corresponding QCD
processes. To obtain radiative dilepton production
[Fig.~\ref{fig1}(a)] by scaling
the corresponding QCD processes [Figs.~\ref{fig1}(b) and (c)], one
simply multiplies the number of virtual gluons from a
quark line by a ratio,
\begin{equation}
        {\cal R}(M^2,Q^2_{\rm max})\equiv
        \frac{d{\cal P}_{q\rightarrow {\rm DL}}}{dM^2}/
        \frac{d{\cal P}_{q\rightarrow {\rm all}}}{dM^2},
                        \label{eq:ratio1}
\end{equation}
for given $M^2$ and $Q^2_{\rm max}$.

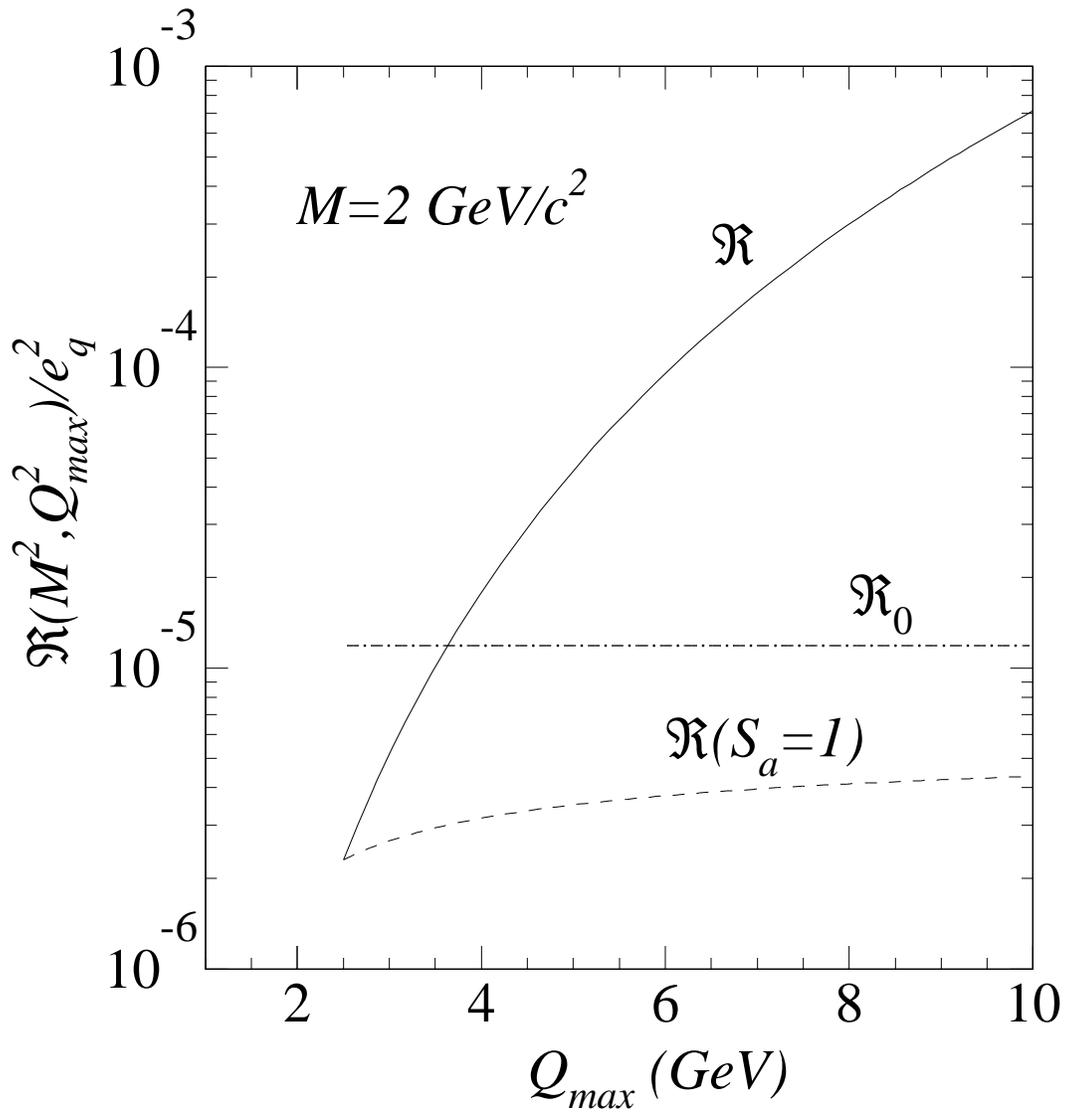
\begin{figure}
\caption{The ratio ${\cal R}(M^2,Q^2_{\rm max})$
        between the probability of
        $q\rightarrow \ell^+\ell^- +q$
        and $q\rightarrow q_i\bar{q}_i +q$, $gg+q$
        processes as functions of
        $Q_{\rm max}$ at fixed $M=2$ GeV/c$^2$, with (solid) and
        without (dashed) the Sudakov form factors. ${\cal R}_0$
        (dot-dashed) is obtained with both Sudakov form factors
        and the $z(1-z)q^2$ dependence of $\alpha_s$ neglected.
        A factor $e_q^2$ is divided out.}
\label{fig2}
\end{figure}

        To demonstrate the effects of the Sudakov form
factor on scaling QCD processes of the Monte Carlo
simulations, we plot in Fig.~\ref{fig2} the ratio
${\cal R}(M^2,Q^2_{\rm max})$ (solid) as a function
of $Q_{\rm max}$ at fixed $M=2$ GeV/c$^2$ together with
the result (dashed) obtained when Sudakov form factors are
set to unity. We also show the value of
${\cal R}_0(M^2,Q^2_{\rm max})$ (dot-dashed) in which
both the Sudakov form factors and $z(1-z)q^2$ dependence
of the coupling constant are neglected. In this case
the ratio is reduced to
\begin{equation}
        {\cal R}_0(M^2,Q^2_{\rm max})=
        \frac{e_q^2}{2\gamma_g(M)}
        \frac{\alpha^2}{\alpha_s(M)^2}
\end{equation}
which is independent of $Q_{\rm max}$, and where
\begin{eqnarray}
        \gamma_g(M)&\equiv&\int_{\epsilon(M)}^{1-\epsilon(M)}dz
                \left[n_fP_{g\rightarrow q\bar{q}}(z)+
                P_{g\rightarrow gg}(z)\right] \nonumber \\
        &=&6\ln[1/\epsilon(M)-1]+9[\epsilon(M)-\frac{1}{2}].
\end{eqnarray}

        Since the Sudakov form factor
takes into account additional branchings preceding
the chosen vertex, it should suppress the probability
distribution of the splitting
$g\rightarrow q\bar{q}, gg$ at the given $M$.
As we see in Fig.~\ref{fig2}, the ratio ${\cal R}(M^2,Q^2_{\rm max})$
is therefore enhanced relative to both
the case when ${\cal S}_a=1$ and to
${\cal R}_0(M^2,Q^2_{\rm max})$ by the inclusion of
Sudakov form factors. The enhancement increases
with $Q_{\rm max}$ as expected due to the increasing
branching probability. Similar to the dilepton fragmentation
functions, the ratio is very sensitive to the scale
$Q_{\rm max}$. We will discuss in Sec.~\ref{sec:iv} how we
choose $Q_{\rm max}$ which is consistent with the Monte Carlo
simulation of QCD cascading. When Sudakov form factors are set to
unity, the dependence of the ratio on $Q_{\rm max}$ only
comes from the $z$ and $q^2$ dependence of the running
strong coupling constant. If both $z(1-z)q^2$ and $z_1(1-z_1)M^2$
in Eq.~\ref{eq:ratio} are replaced by $M^2$, the ratio
${\cal R}_0$ becomes larger and is independent of $Q_{\rm max}$
as shown in Fig.~\ref{fig2}.

\section{SIMULATION OF DIRECT DY PROCESSES}

It is relatively easier to simulate the lowest order
process of direct Drell-Yan by multiplying the differential
cross section of $q\bar{q}\rightarrow q_i\bar{q_i}, gg$
by the ratio
\begin{equation}
  {\cal R}_{\rm DY}=\frac{d\sigma_{q\bar{q}\rightarrow DY}}
  {\sum_{i=1}^{n_f}d\sigma_{q\bar{q}\rightarrow q_i\bar{q}_i}
    +d\sigma_{q\bar{q}\rightarrow gg}}.
\end{equation}
It is more subtle, however, to include  QCD corrections.
These higher order corrections which give rise to a so-called
$K$-factor have been studied extensively \cite{DYK1}. The problem here
is how to include this $K$-factor in scaling QCD processes
to obtain the DY cross sections.

The first order correction to DY process in QCD comes from
the ``annihilation'' $q\bar{q}\rightarrow g+ {\rm DY}$, the
``Compton'' $q+g\rightarrow q+{\rm DY}$ process and the
virtual corrections \cite{DYK1}.  Like in deeply inelastic lepton
nucleon scatterings (DIS), there are infrared singular
and finite contributions from these corrections. The infrared
singular and part of the finite terms can be absorbed into
the quark and antiquark distribution functions which are
defined in DIS processes and should be evaluated at the
scale of $M^2$ according to Altarelli-Parisi evolution equations \cite{AP}.
What are left over are finite and scheme-independent
contributions from the higher order corrections. One
can find detailed discussions in, {\it e.g.}, Ref.\cite{FIED}.
What we want to point out here is that the dominant
contribution to the $K$-factor of about 2 in the ($p_T$-integrated)
mass spectrum
of the direct DY is from the virtual corrections. The contributions
from real corrections which depend on both the quark and gluon
distribution functions are relatively very small.
Therefore, as a first approximation, we
can include higher order corrections to
direct DY process in our simulation by multiplying
the QCD cross sections of $q\bar{q}\rightarrow q_i\bar{q}_i, gg$
by an effective $K=2$ factor.  This $K$ factor in principle
now includes
both real and virtual corrections.
The quark distribution functions in the cross section
should be evolved and evaluated at scale $M^2$.

In the Monte Carlo simulations, one could also include
the real QCD-corrections explicitly
to DY process by counting the number
of similar QCD processes, $q\bar{q}\rightarrow gg$,
$q+g\rightarrow q+g$ and scaling them
by some calculable
ratio as has been done in  Ref.\cite{KGJK}. However,
one still has to include the virtual
corrections which can be characterized as an effective
multiplicative factor, but which now
differs from the normal overall DY $K$-factor.
This is exactly the problem one has to face if one wants
to simulate the $p_T$ distribution of DY dilepton
pairs, whose large $p_T$ tail mainly comes from
real QCD-corrections. The lowest order DY process
only contributes to the small $p_T$ part of the
spectrum by including the intrinsic $p_T$ of quarks and
anti-quarks. The virtual corrections to the lowest
order DY can be taken into account by using an effective
`$K$'-factor. However, one must be very careful not to
include the real corrections in this effective `$K$'-factor.

         In the Monte Carlo simulations of QCD processes
in  hadronic collisions, one usually also uses an
effective $K$-factor to take into account higher
order corrections \cite{PYTH}. However, this $K$-factor should not be
included when one scales the number of
$q\bar{q}\rightarrow gg$ and $q+g\rightarrow q+g$
processes by some ratio to calculate the real
QCD-corrections to DY process. Otherwise,
double-counting may occur.

\section{NUMERICAL RESULTS}
\label{sec:iv}

        To perform the Monte Carlo simulations, we use
PYTHIA \cite{PYTH} subroutines for QCD hard scatterings and
the associated bremsstrahlungs as adapted in HIJING
model \cite{HIJING}. HIJING is a Monte Carlo model developed
for parton and particle production in high energy $pp$, $pA$ and
$AA$ collisions. In this model, multiple minijet production
at $NN$ level is calculated in the eikonal formalism \cite{WANG91}.
As in many other models which attempt to merge low and high
$p_T$ dynamics, a $p_T$ cutoff scale $p_0$ has to be
introduced, which will limit the invariant masses of
produced dileptons in our simulation. For nuclear
interactions, binary approximation is assumed for independent
hard scatterings. Jet quenching due to final state
interaction of produced partons with string-like soft
mean-field was also included in the original HIJING model \cite{WANG92}.
We switch off these final state interactions to simplify
our study here so that we can compare the numerical
results with semi-analytical calculations.
The initial parton distribution functions
of a nucleon is taken to be Duke-Owens parameterization \cite{DO84}
set 1. Nuclear shadowing and its scale dependence are
also taken into account as in Ref.~\cite{ESKO92}.
Impact parameter dependence of the nuclear
parton distributions is modeled in
according to Refs.\cite{HIJING,ESKO91}. However, at
$\sqrt{s}=200$ AGeV, the shadowing effects on the
associated dilepton production are small
as seen in Ref. \cite{KEXW}.

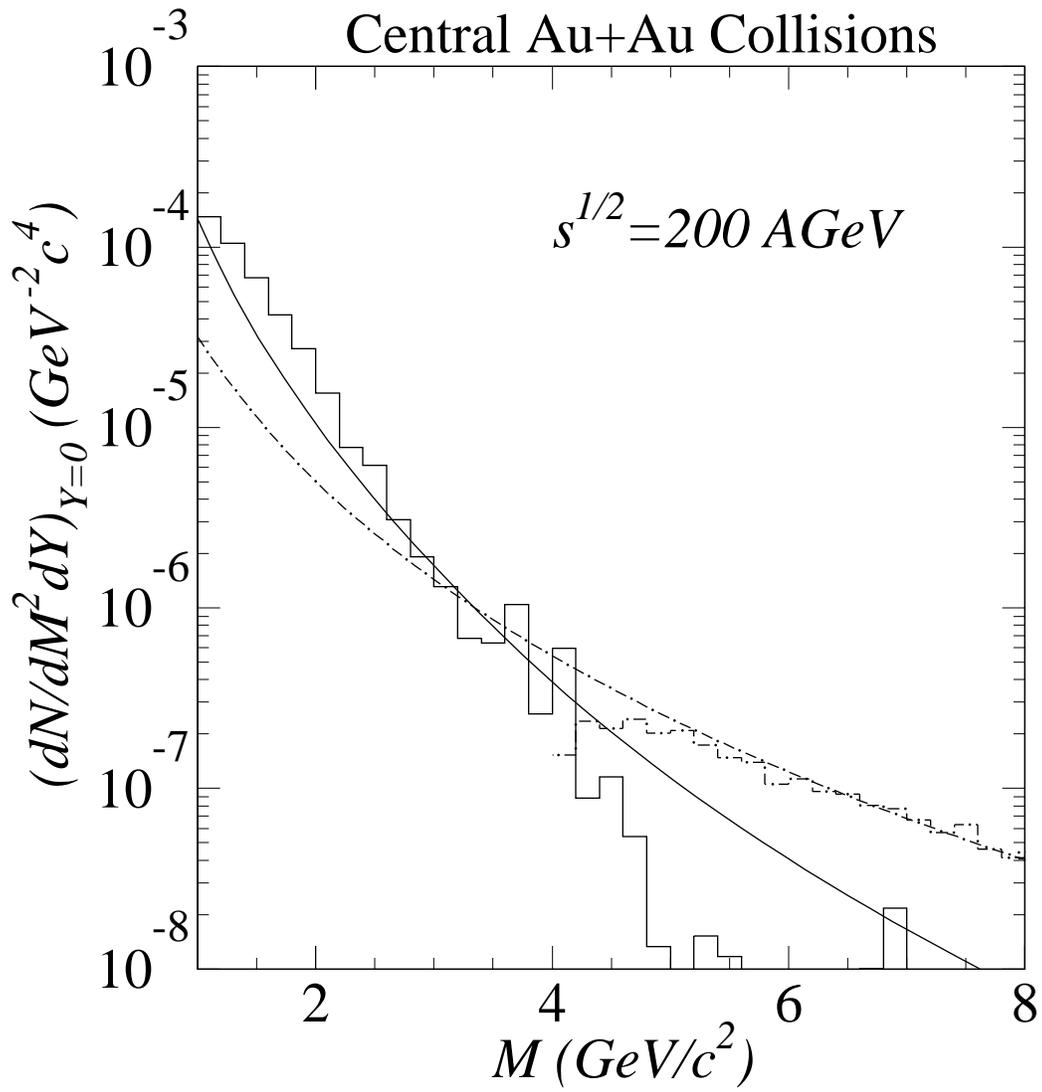
\begin{figure}
\caption{Mass spectrum of the minijet-associated
  (solid histogram) and direct DY (dashed histogram)
  dileptons from the Monte Carlo simulation and our
  direct calculation (solid and dashed curves, respectively)
 (with $Q_{\rm max}=2p_T$, $\Lambda=0.4$ GeV and $\mu_0 = 0.5$ GeV)
 in central $Au+Au$ collisions at $\protect\sqrt{s}=200$ AGeV.}
\label{fig3}
\end{figure}

        During the final state radiation, we count the
number of virtual gluons with given invariant mass
$M$ which are radiated from quark lines. The
maximum virtuality $Q_{\rm max}$
of the quark should be the invariant mass of its
parent parton minus the minimum virtuality $\mu_0$ of
its sister parton. If there is no bremsstrahlung prior to
this branching vertex, $Q^2_{\rm max}$ should be related
to the transverse momentum transfer $p_T$
of the corresponding hard scattering. In order to conserve
both energy and momentum, the two produced partons from a
hard scattering are combined together in PYTHIA \cite{PYTH2}
to initiate final state radiation. The total virtuality of the
two parton system is chosen to be $2p_T$. Since both
of the partons must have at least a virtuality of $\mu_0$,
the maximum virtuality of the selected quark immediately
after the hard scattering should be $Q_{\rm max}=2p_T-\mu_0$.
With given $M^2$ and $Q^2_{\rm max}$, we then can calculate
the number of dileptons produced from the final state
radiation by multiplying the number of these
radiated virtual gluons with the ratio
${\cal R}(M^2,Q^2_{\rm max})$. Shown by the
solid histogram in Fig.~\ref{fig3}, is the invariant
mass distribution of the radiated dileptons thus
obtained for central $Au+Au$ collisions at RHIC
energy. In the simulation, the parton shower is
terminated whenever a minimum virtuality
$\mu_0=0.5$ GeV is reached. Then the parton is put
on shell and considered real. So the minimum
invariant mass of our selected virtual gluons, thus
also of the dileptons, is $2\mu_0=1$ GeV, according to
Eq.~\ref{eq:eps}.

        We also plot in Fig.~\ref{fig3} our
semi-analytical calculation of the radiative
dilepton production (solid curve) which
agrees quite well with the simulated result.
The small differences both in the total number
and the slope of the distribution
could come from several simplifications we made
in our semi-analytical approach. As stated in
Ref. \cite{KEXW},  we did not fully take
into account the kinematic restrictions (Eqs.~\ref{eq:eps},
\ref{eq:e12}) at every stage of the radiation tree
in the calculation of the dilepton fragmentation
functions. The variable in the strong coupling
constant is taken to be $q^2$ instead of the
relative transverse momentum $q_T^2\approx z(1-z)q^2$.
The fragmentation function approach has only
one branching tree corresponding to a simple
ladder structure, whereas the Monte Carlo
simulation takes into account all possible
branching trees, thereby enhancing the small $M$
dilepton production.  In order to be as consistent
as possible in both calculations in Fig.~\ref{fig3},
we have chosen the same scales $Q_{\rm max}=2p_T$ and
$\Lambda=0.4$ GeV in the dilepton
fragmentation functions as have been used in
the Monte Carlo simulation \cite{PYTH}.

To simulate the lowest order direct DY process, we
simply count the number of similar QCD subprocesses,
$q\bar{q}\rightarrow q\bar{q}$, $gg$ at fixed
$\hat{s}=M^2$. We then multiply the number by
the ratio ${\cal R}_{\rm DY}$ to obtain the
number of direct DY dileptons, which is shown
as the dashed histogram in Fig.~\ref{fig3}. We also
compare the result to the parton model calculation
(dashed curve) with the same set of parton distribution
functions as used in the simulation.  Higher order
corrections are included by multiplying a $K=2$ factor
in both the simulation and analytical calculation.
In terms of $p_T$ and rapidities $y_{1,2}$ of minijets,
the invariant mass of the dilepton is
\begin{equation}
  M^2=2p_T^2[1+\cosh(y_1-y_2)].
\end{equation}
Since we have a cutoff $p_0=2$ GeV/c for $p_T$, the lower
limit of the dilepton mass from the Monte Carlo simulation
is then $M\leq 4$ GeV/c$^2$. The analytical
calculation can go as low as the initial
scale of the structure functions, $Q_0=2$ GeV.

\section{CONCLUSIONS}

        In this paper, we have studied minijet-associated
dilepton production in ultra-relativistic nuclear collisions
through Monte Carlo simulations to check our previous
semi-analytical calculation \cite{KEXW} through the fragmentation
function approach.  We derived a formula for the differential
ratio by which we multiply the number of similar QCD
processes to obtain dilepton production from the Monte
Carlo simulation of QCD cascading. Using this ratio,
we found that our semi-analytical calculation is
consistent with Monte Carlo simulations. The difference
between the two due to some simplifications we
made in the fragmentation function approach is
small.

Most importantly, we found that Sudakov form factors
which were not included in the ratio in Ref.~\cite{KGJK}
are essential for us to give the right
results. If neglected, the resultant dilepton rate
from final state radiation would differ from our
early semi-analytical calculation by orders of magnitude.
Due to the same reason, the differential ratio is
quite sensitive to the maximum virtuality $Q_{\rm max}$ of
the branching processes, similar to the fragmentation
function approach. One therefore has to choose its
value to be consistent with what is used in the Monte
Carlo simulation of QCD cascading.

We also simulated the direct Drell-Yan processes of
dilepton production and compared it to semi-analytical
calculation in the parton model. We pointed out the complication
in including higher order corrections in the Monte Carlo
simulations and the possibility of double counting. We
believe this is especially important when one wants to
simulate dilepton production through the final state
parton rescatterings \cite{KGJK,RANF} in a dense partonic system like
a quark-gluon plasma. Unlike in hadronic scatterings
where infrared singularities due to real and virtual
corrections can be absorbed into the definition
of QCD evolved parton distribution functions, the
screening mass due to resummation of hot thermal
loops \cite{BP} naturally  regulates
the infrared divergences. However,
one still has important contributions from both
real and virtual corrections \cite{RUSS}. In order to take into
account these corrections in a Monte Carlo simulation,
one needs to analyze the higher order calculation
in finite temperature QCD in detail.

\acknowledgments
This work was supported by the Director, Office of Energy
Research, Division of Nuclear Physics of the Office of High
Energy and Nuclear Physics of the U.S. Department of Energy
under Contract No. DE-AC03-76SF00098.
KJE thanks Magnus Ehrnrooth foundation, Oskar \"{O}flund
foundation, and Suomen Kulttuurirahasto for partial financial
support.

\end{document}